\documentclass[fullpage,11pt]{article}

\title{Initial Conditions of Closed Classical Orbits from Quantum Spectra}
\author{
Michael Courtney
\\
\small \it Department of Physics, Massachusetts Institute of Technology,\\
\small \it 77 Massachusetts Avenue, Cambridge, Massachusetts 02139
}
\date{\small Chaos {\bf 6}, 63 (1996); DOI:10.1063/1.166157}

\begin{document}
\maketitle
\begin{abstract}
A method is presented for determining the initial conditions of
classical orbits from the quantum spectra of the diamagnetic
hydrogen atom.
Each classical trajectory which is closed at the nucleus
produces a sinusoidal fluctuation in the photoabsorption
spectrum.  The amplitude
of each orbit's contribution appears
in the Fourier transform of a spectrum computed at constant
scaled energy.  For a given initial state,
closed-orbit theory gives the dependence
of this recurrence amplitude on the initial angle of an orbit.
By comparing the recurrence amplitudes
for different initial states, the initial conditions of closed
classical orbits are determined from quantum spectra.

\end{abstract} 

\section{Introduction}
According to the Correspondence Principle, the solutions to
Schr\"{o}dinger's equation for a given system should contain
the classical motion of the same system as a subset 
(the ``classical limit").
In practice, however,
gaining information about the classical dynamics of a system
from the quantum behavior has proved challenging, particularly
for chaotic systems.
In some cases, the distribution of eigenvalues can be used
to discern whether the classical motion is regular or
chaotic \cite{review,delande}.  Spectral periodicities are related to
periodic orbits \cite{gutzwiller,welge} and can be used to
determine such classical
quantities as the period, action, and stability of the
periodic orbits \cite{review,wintgen}.
This paper shows that
spectral periodicities of the diamagnetic hydrogen atom
can be used to determine the initial conditions of the
periodic orbits which pass through the 
origin (closed orbits).

Semiclassical quantization techniques reverse the causal
role between quantum and classical behavior by using the
classical solutions to construct approximate quantum solutions.
EBKM tori quantization \cite{gutzwiller} gives
approximate eigenvalues of conservative integrable systems.
Periodic-orbit theory predicts the level density from
a sum over periodic orbits and is applicable
in regimes of integrable and 
non-integrable classical motion \cite{gutzwiller}.
The closely related closed-orbit theory \cite{du}
predicts photoabsorption spectra from a sum over
the orbits which are closed at the nucleus.

Periodic-orbit theory can also be used to glean classical information
from quantum level densities \cite{wintgen}.
The main result of periodic-orbit theory \cite{gutzwiller} is
an expression for the fluctuating part of the level density,
\begin{equation}
\label{eqn:trace}
g_c(E) = \sum_k \sum_{n=1}^{\infty} A_{\it nk}
e^{i(nS_k - \alpha_{\it nk} \pi/2)},
\end{equation}
(atomic units).
The index $k$ distinguishes the primitive periodic orbits: the shortest
period orbits for a given set of initial conditions.
$S_k$ is the action of a periodic orbit.
Each primitive orbit retraces itself, leading to new orbits
with action $nS_k$, where $n$ is an integer.
Hence, every repetition
of a periodic orbit is another periodic orbit.
The quantity $A_{\it nk}$ is related to the
stability of an orbit,
and $\alpha_{\it nk}$ is the orbit's Maslov index.

\section{Closed Orbit Theory in Diamagnetic Hydrogen}
The Hamiltonian of diamagnetic hydrogen is
\begin{equation}
H = \frac{p^2}{2} - \frac{1}{r} + \frac{1}{8}B^2\rho^2,
\end{equation}
where the magnetic field $B$ is taken to be along the
$z$ axis.  This Hamiltonian can be scaled 
so that for spectra at constant scaled energy, $\epsilon = EB^{-2/3}$,
the fluctuating part of the level density is \cite{wintgen}
\begin{equation}
\tilde{g}_c(w) = \sum_k \sum_{n=1}^{\infty} \tilde{A}_{\it nk}
e^{i(2\pi n w\tilde{S_k} - \alpha_{\it nk} \pi/2)},
\end{equation}
where $w = B^{-1/3}$, and $\tilde{S} = S/(2\pi w)$ is the scaled action.
At fixed $\epsilon$,
the classical dynamics has no dependence on $w$.  Consequently,
the Fourier transform of an energy level spectrum computed at
fixed $\epsilon$ as a function of $w$ gives peaks which
lie at the scaled action of periodic orbits and
whose heights are related to the stability of the 
orbits \cite{scaling}.

Closed-orbit theory \cite{du} is similar to
periodic-orbit theory, except that closed-orbit theory
is applicable only to atomic and molecular spectra and
yields the oscillator strength density from a specified
initial state whereas periodic-orbit theory yields
the density of states.
Only orbits that begin and end at the nucleus are important
in closed-orbit theory.
Physically, these are
associated with the outgoing waves that are generated when a tightly bound
electron is excited to a high-lying state.  For diamagnetic hydrogen,
every orbit which is closed at the nucleus is also a periodic
orbit whose period is equal to either the closure time or twice
the closure time.

According to closed-orbit theory, the average oscillator strength
density at constant $\epsilon$ is given by a smooth background
plus an oscillatory sum of the form \cite{du}
\begin{equation}
f(w) = \sum_k \sum_{n=1}^{\infty} D^{i}_{\it nk}
\sin(2\pi nw\tilde{S_k} - \phi_{\it nk}).
\end{equation}
$\phi_{nk}$ is a phase that 
depends on the Maslov index and other details of the orbit.
${D}^i_{\it nk}$ is the recurrence amplitude of
a closed orbit for a given initial state (labeled $i$). 
It contains information about the
stability of the orbit, its initial and final directions,
and the matrix element of the dipole
operator between the initial state and a zero-energy
Coulomb wave.
The Fourier transform of an oscillator strength spectrum computed at
fixed $\epsilon$ as a function of $w$ is called a recurrence
spectrum, because it gives peaks which
correspond to the scaled action of closed orbits and
whose heights correspond to ${D}^i_{\it nk}$.

\section{Method for Determining Initial Angles}
The dependence of the recurrence amplitude ${D}^i_{\it nk}$
on the initial and final angles of an orbit can be used to
determine these angles from computed photoabsorption spectra.
We can write the recurrence amplitude as
\begin{equation}
\label{eqn:cot}
{D}^i_{\it nk} = F^i(\theta_i,\theta_f) G_{\it nk},
\end{equation}
where $\theta_i$ and $\theta_f$ are the initial and final
angles the orbit makes with the $z$ axis at the origin.
(For a specified scaled energy, the initial conditions
of a closed orbit are completely determined by
its initial angle.)
The dependence on the initial and final angles of the orbit
and on the initial state is
completely contained in the function $F^i(\theta_i, \theta_f)$.
$G_{\it nk}$ is related to the stability of the 
orbit \cite{du}.
Several factors simplify 
the discussion for the scaled energy ($\epsilon = -0.7$) 
considered here \cite{simplify}.
The dynamics is near integrable in this regime
and all closed orbits have either $\theta_f = \theta_i$
or $\theta_f = \pi - \theta_i$.  Furthermore,
$F^i(\theta_i, \theta_f) = F^i(\theta_i, \pi - \theta_f)$.
Consequently, the angular dependence can be considered
a function of a single angle, $\theta$.

The form of $F(\theta)$ for a given initial state
can be obtained from Du and Delos \cite{du}.
For $m = 0$ final states,
\begin{eqnarray}
F^{\it 3s}(\theta) & = & 0.625 \sin{\theta}[P_1(cos{\theta})]^2  \\
F^{\it 2p}(\theta) & = &0.5797 \sin{\theta}
[P_2(cos{\theta}) + P_0(cos{\theta})/4]^2 \\
F^{\it 3d}(\theta) & = & 0.8505 \sin{\theta}
[P_3(cos{\theta}) + 2P_1(cos{\theta})/9]^2,
\end{eqnarray}
where the $P_l(\cos{\theta})$ are Legendre polynomials.
(For orbits parallel to the field, there is
no $\sin{\theta}$ term.)
The angular functions are normalized \cite{normalized}
such that
\begin{equation}
\int_0^\pi F^i(\theta)\, d\theta = 1.
\end{equation}
Since all the dependence on the initial state is contained
in $F^i(\theta)$,
\begin{equation}
\frac{D^i_{\it nk}}{D^j_{\it nk}} = \frac{F^i(\theta)}{F^j(\theta)}.
\end{equation}
Consequently, the ratio ${F^i(\theta)}/{F^j(\theta)}$
for a particular closed orbit can be determined by dividing
the recurrence amplitudes for the corresponding peak
in the computed
recurrence spectrum.

\begin{figure}
\vspace{2.0in}
\label{fig:ratio}{\small
Fig. 1: Predicted ratios of recurrence amplitudes
from different initial states.}
\includegraphics{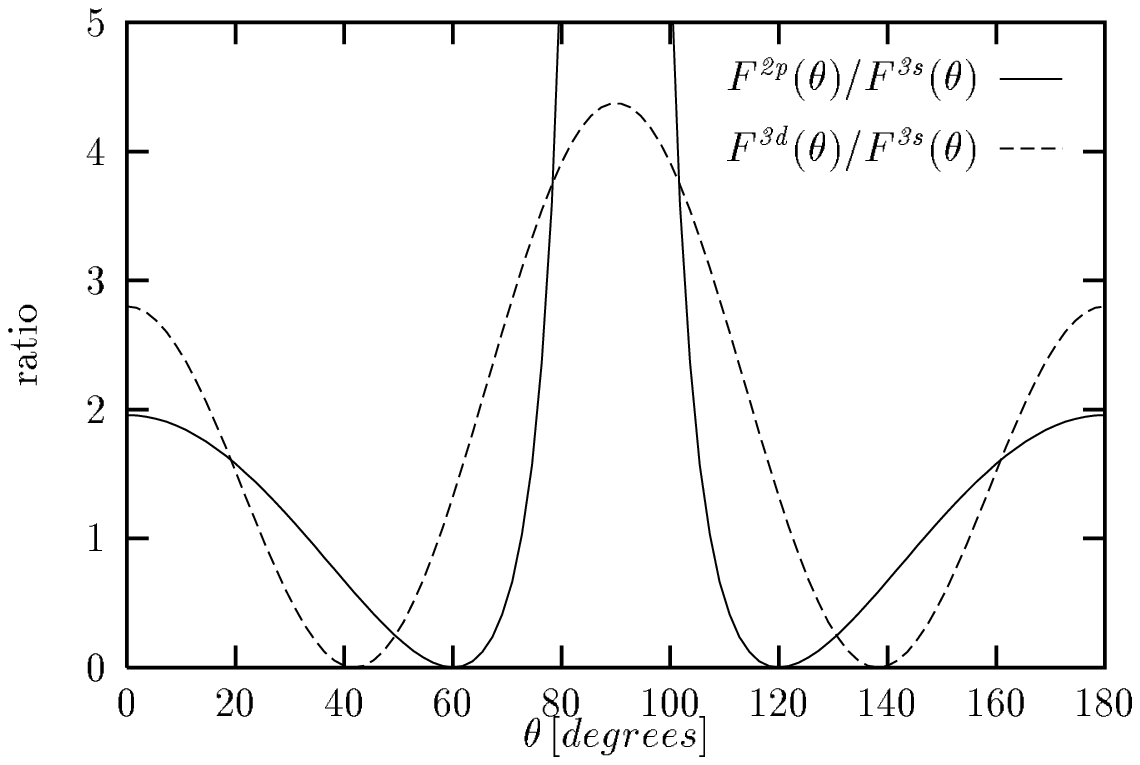}
\end{figure}

Fig. 1 shows 
$F^{\it 2p}(\theta)/F^{\it 3s}(\theta)$ and
$F^{\it 3d}(\theta)/F^{\it 3s}(\theta)$.  These ratios
are symmetric about $90^\circ$.  A given ratio corresponds
to either two or four possible initial angles, but since
a closed orbit of diamagnetic hydrogen with initial angle
$\theta$ has a symmetric partner of the same shape with
initial angle $\pi - \theta$, only the range
from $0 \le \theta \le \pi/2$ is of interest.
This leaves either one or two possible initial angles, depending 
on the value of the ratio, but
this ambiguity can be resolved by looking at both ratios.

\section{Recurrence Spectra}
We now turn to the recurrence spectra.  A number of methods
are available for computing spectra of diamagnetic
hydrogen \cite{delande,wintgen2,clark}.
Diagonalizing the 
Hamiltonian matrix in a spherical hydrogenic basis
is adequate for $\epsilon = -0.7$ \cite{zimmerman}.
The oscillator strength spectrum, $f(w)$,
is multiplied by $w^3$ to remove the global variation
of oscillator strength over the range of fields ($31 < w < 133$)
and normalized
so that
\begin{equation}
\int_{w_{min}}^{w_{max}} w^3 f(w)\, dw = 1.
\end{equation}
This produces narrower peaks in the recurrence spectrum and 
allows for comparison with recurrence spectra of different initial
states.  (In principal, one can obtain arbitrarily high resolution in the
recurrence spectrum by computing spectra to very large $w$.
This allows identification of orbits arbitrarily close in action.)

\begin{figure}
\vspace{2.0in}
\label{fig:3srs}\small{
Fig. 2: Recurrence amplitude for odd-parity, $m = 0$ final
states at $\epsilon = -0.7$ excited from the $3s$
initial state.  (The recurrence amplitude is the
square root of the power spectrum.)}
\includegraphics{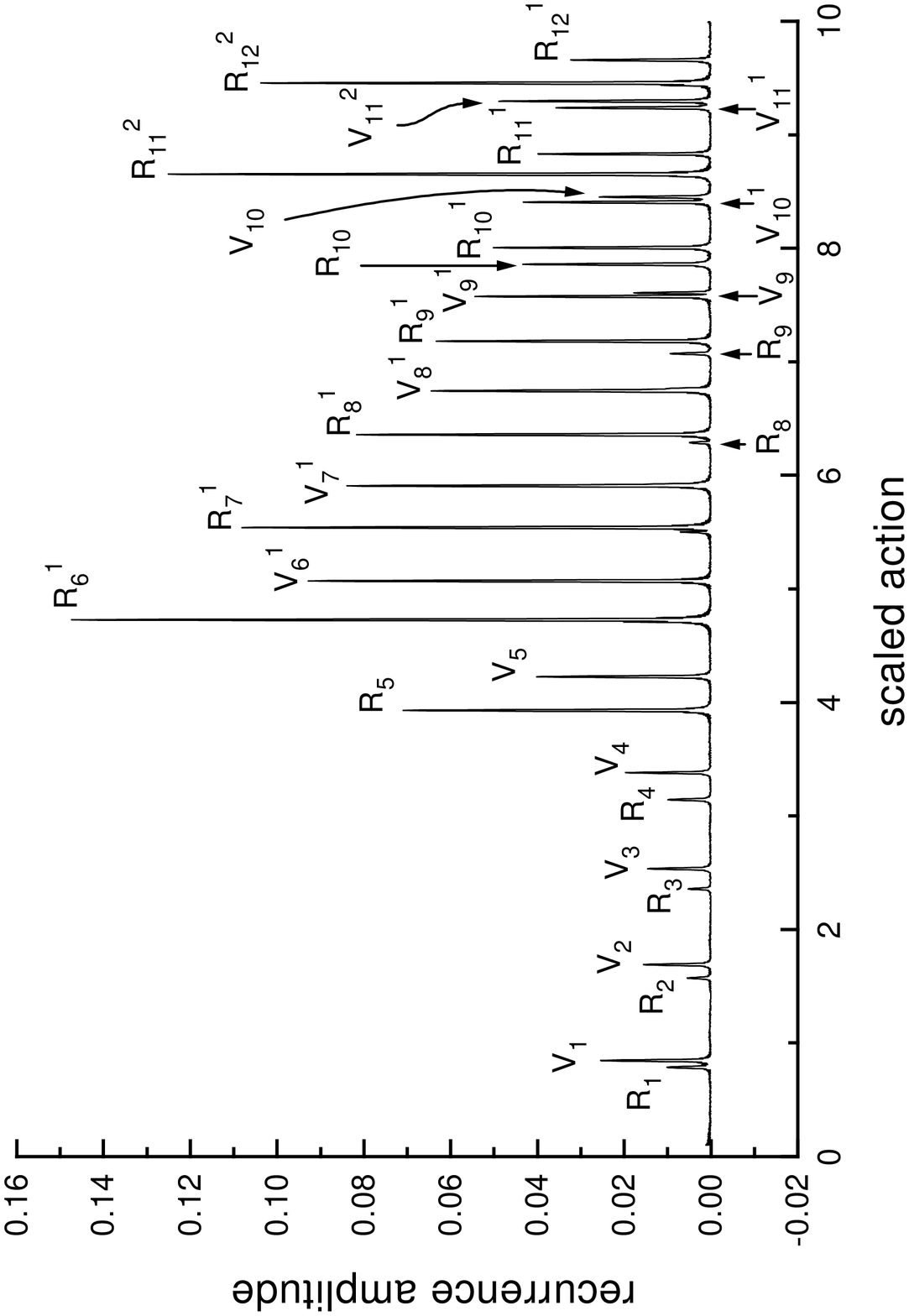}
\end{figure}

The $m = 0$ odd-parity recurrence spectrum for the $3s$ initial
state is shown in Fig. 2.
(van der Veldt {\it et al.} have analyzed a similar
experimental recurrence spectrum for the $2s$ initial
state in helium \cite{van}.)
To interpret the spectrum, it is helpful to
recall the basic structure of closed orbits in
diamagnetic hydrogen.
At large negative scaled energies, there are three primitive
short-period orbits:  one moves on the $\rho$ axis, and
two move on the $\pm z$ axis \cite{review}.
Most longer-period primitive closed orbits are
created by bifurcations of these orbits
and their repetitions.

The orbits on the $\rho$ axis and those which bifurcate
from them are called {\it rotators}.
Peaks corresponding to these orbits will be
labeled $R_n^b$, where $n$ denotes the
repetition of the orbit on the $\rho$ axis from which
the orbit bifurcated, and $b$ distinguishes
between different orbits which bifurcated from the same
parent.
The orbits on the $z$ axis and those which bifurcated
from them are called {\it vibrators}.
Recurrences corresponding to these orbits are labeled
$V_n^b$, where $n$ and $b$ have the same meaning as for
the rotators.
For $\tilde{S} < 4$ the recurrence spectrum is dominated by
repetitions of the perpendicular and parallel
orbits (labeled $R_n$ and $V_n$, respectively) \cite{shaw2}.
For $\tilde{S} > 4$ other orbits are present which were
born by bifurcations of these orbits,
and the recurrence spectrum becomes more complicated.

\begin{figure}
\vspace{2.5in}
\label{fig:mirror}{\small
Fig. 3  Top: Recurrence amplitude for even-parity, $m = 0$
final states at $\epsilon = -0.7$ excited from the
$2p$ initial state.  Bottom: Recurrence amplitude for
odd-parity, $m = 0$ final
states at $\epsilon = -0.7$ excited from the $3s$
initial state.  This recurrence spectrum is multiplied
by $-1$ for comparison.}
\includegraphics{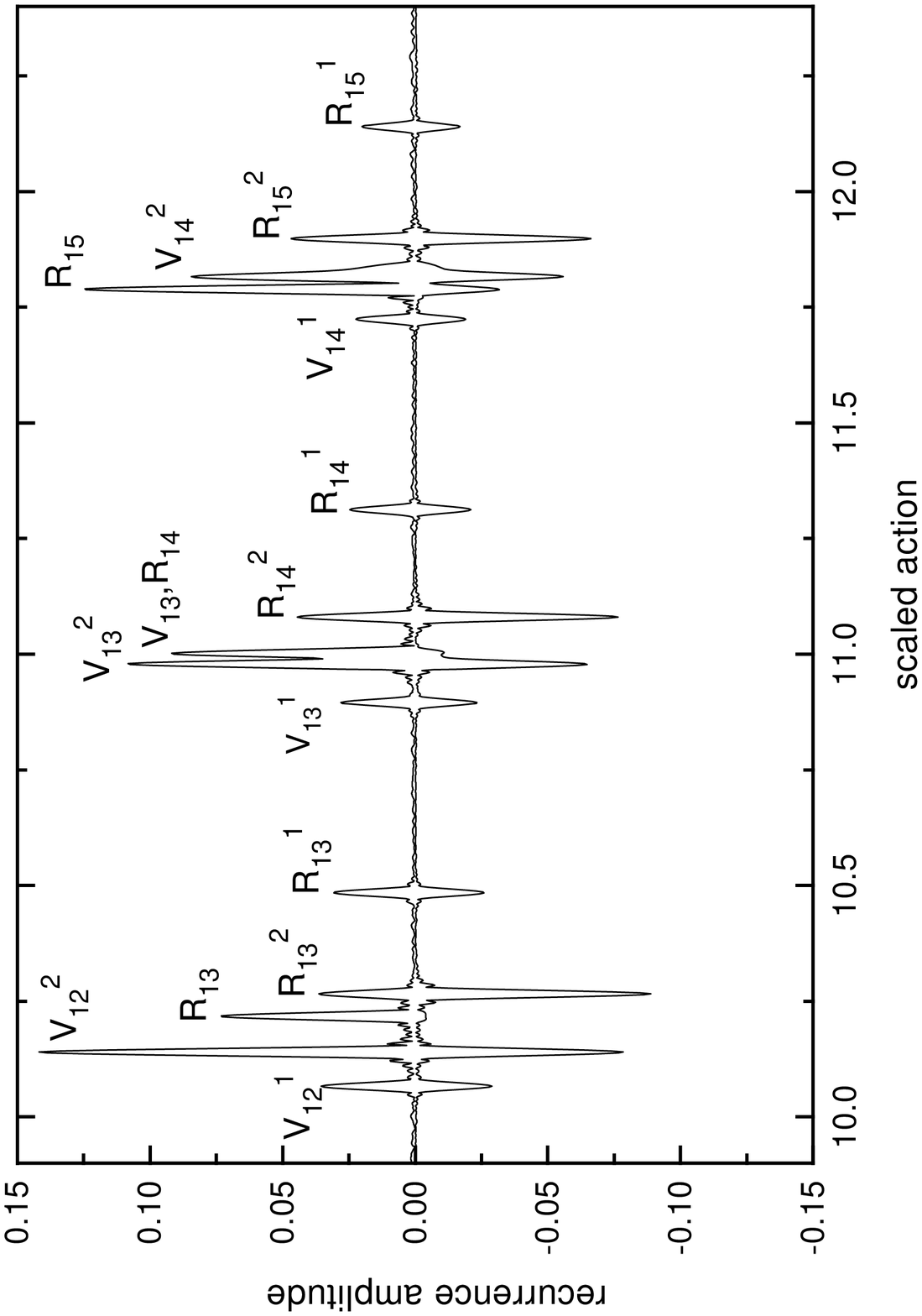}
\end{figure}

\section{Determination of Initial Angles}
This method for determining initial angles can be tested by
considering recurrence spectra for $9.9 < \tilde{S} < 12.4$,
which are shown in Fig. 3 for the $2p$ and $3s$
initial states.  This range of action has a number of
orbits with non-trivial (not $0^\circ$ or $90^\circ$) initial
angles and is sufficiently low that the density of
orbits allows most recurrences to be resolved.

The method for determining the initial angle of an orbit
is demonstrated by considering the peak at $\tilde{S} = 10.0661$.
The amplitudes for the $2p$ and $3s$ initial states are
$D^{\it 2p} = 0.0361$ and $D^{\it 3s} = 0.0294$.
Their ratio corresponds to
$F^{\it 2p}(\theta)/F^{\it 3s}(\theta) = 1.228$.
Two initial angles give this ratio: $\theta = 28.51^\circ$ and
$\theta = 73.47^\circ$.  
This ambiguity can be resolved
by considering the amplitude for the $3d$ initial
state.  
$D^{\it 3d} = 0.0181$ which corresponds to
$F^{\it 3d}(\theta)/F^{\it 3s}(\theta) = 0.615$.
Once again, two initial angles give this ratio:
$\theta = 29.07^\circ$ and $\theta = 53.90^\circ$.
Consequently, the initial angle is determined
to be near $\theta = 28.51^\circ$.  This is in good
agreement with the actual initial angle, 
$\theta^{\it cl} = 29.10^\circ$.

\begin{table}
\caption{\small Predicted initial angles from two different amplitude
ratios are compared with actual classical initial angle, 
$\theta^{\it cl}$.  All angles are in degrees.
}
\label{tab:angles}
\begin{tabular}{ccccccc}
$\tilde{S}$ & $\frac{D^{\it 2p}}{D^{\it 3s}}$ & $\theta^{2p/3s}$ & $\frac{D^{\it 3d}}{D^{\it
3s}}$ & $\theta^{3d/3s}$ & $\theta^{\it cl}$ & orbit\\ \hline
10.066 & 1.228 & 28.51 & 0.615 & 29.07 & 29.10 & $V_{12}^1$ \\ 
10.140 & 1.815 & 12.04 & 2.209 & 12.88 & 14.23 & $V_{12}^2$ \\ 
10.266 & 0.406 & 45.64 & 0.086 & 46.28 & 45.35 & $R_{13}^2$ \\
10.485 & 1.181 & 29.54 & 0.481 & 30.68 & 31.00 & $R_{13}^1$ \\
10.895 & 1.195 & 29.23 & 0.557 & 29.74 & 29.47 & $V_{13}^1$ \\
10.979 & 1.671 & 17.27 & 1.758 & 17.69 & 18.05 & $V_{13}^2$ \\
11.080 & 0.583 & 41.85 & 0.012 & 40.14 & 41.84 & $R_{14}^2$ \\
11.312 & 1.172 & 29.73 & 0.488 & 30.58 & 30.71 & $R_{14}^1$ \\
11.724 & 1.182 & 29.52 & 0.577 & 29.52 & 29.71 & $V_{14}^1$ \\
11.816 & 1.487 & 22.44 & 1.406 & 21.09 & 20.71 & $V_{14}^2$ \\
11.898 & 0.707 & 39.29 & 0.043 & 38.62 & 39.31 & $R_{15}^2$ \\
12.140 & 1.191 & 29.31 & 0.495 & 30.49 & 30.52 & $R_{15}^1$

\end{tabular}
\end{table}

This method can be used to determine the initial angle of
the closed orbits corresponding to many of the peaks
in Fig. 3.  Table \ref{tab:angles} shows
that this method gives agreement with the initial angles
obtained classically.  
Some of the closed orbits corresponding to peaks in
Fig. 3 are shown in Fig. 4.

The initial angles with the
largest errors are for orbits where one or more of
the recurrence amplitudes is small or where the ratio
$F^{\it i}(\theta)/F^{\it j}(\theta)$ has a small slope.
If the recurrence amplitudes themselves are small, there
is a larger relative error in the computed amplitudes
due to noise in the Fourier transform.  There is also
a larger relative error in the semiclassical approximations
for small recurrences, because the semiclassical approximations
used to derive Eq. \ref{eqn:cot}
neglect the contribution of a closed orbit's neighbors to
the angular dependence of its recurrence strength \cite{shaw2}.
The error in the predicted initial angle is large if
the ratio of angular functions has a small slope because
a small error in the computed recurrence amplitudes
or in the semiclassical angular functions leads to
a large error in angle.
In principle, the errors in the semiclassical approximations
become smaller as $\hbar \rightarrow 0$.  This limit can
be approached practically in diamagnetic hydrogen by
looking at smaller magnetic fields at a given scaled
energy.  The difficulty of small slopes in the angular
ratios can be remedied by considering a range
of initial states so that, for every angle, at least one of the
angular ratios has a large slope.
A weighted average over
many initial states should produce more accurate initial angles
than consideration of a few initial states.

\begin{figure}
\vspace{0.7in}
\label{fig:co}\small{
Fig. 4: Some closed orbits corresponding to peaks in Fig. 3
and Table \ref{tab:angles}.  
(a) $V_{14}^1$;
(b) $V_{14}^2$;
(c) $R_{15}^2$;
(d) $R_{15}^1$.}
\includegraphics{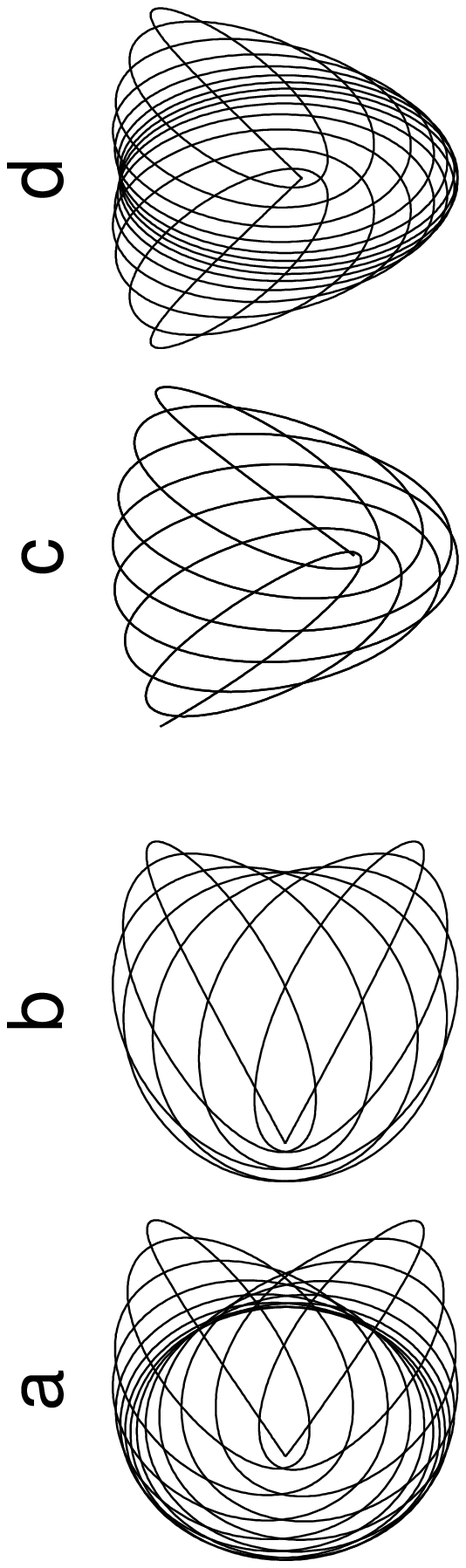}
\end{figure}

\section{Conclusion}
In summary,
a method is presented for determining the initial conditions of
classical orbits from the quantum spectra of the diamagnetic
hydrogen atom.
However, this quantum mechanical method is an inefficient 
choice for searching for closed classical orbits.  More
efficient classical methods are
available for finding closed and periodic orbits by directly integrating
Hamilton's equations \cite{baranger}.
This development of a quantum mechanical method of finding closed
orbits is a practical step toward
understanding how the solutions to Schr\"{o}dinger's equation yield the
solutions to Hamilton's equations in the classical limit.
It has been shown to be accurate in the near-integrable
regime of diamagnetic hydrogen.  A simple generalization
should be applicable in more strongly chaotic regimes.

\section{Acknowledgements}
I am grateful for valuable discussions with
John Delos, John Shaw, and Dan Kleppner.
I also thank Dominique Delande
for allowing the use of his implementation of the Lanczos
algorithm for matrix diagonalization.


\begin{thebibliography}{99} 

\bibitem{review} H. Hasegawa, M. Robnik, and G. Wunner, No. 98, New
Trends in Chaotic Dynamics of Hamiltonian Systems, Progress in
Theoretical Physics, Supplement, pp. 198 (1989);
H. Friedrich and D. Wintgen, Phys. Rep. {\bf 183}, 38 (1989).

\bibitem{delande}
D. Delande and J. C. Gay, Phys. Rev. Lett.
{\bf  57}, 2006 (1986).

\bibitem{gutzwiller} {\it Chaos in Classical and Quantum
Mechanics}, M. C. Gutzwiller, Springer-Verlag (1990).

\bibitem{welge} A. Holle, J. Main, G. Wiebusch,
H. Rottke, and K. H. Welge, Phys.  Rev.  Lett.  {\bf 61},
161 (1988).

\bibitem{wintgen} D. Wintgen, Phys. Rev. Lett.
{\bf 58}, 1589 (1987).

\bibitem{du}
M.L. Du and J.B. Delos, Phys. Rev. A {\bf  38}, 1896 (1988);
{\bf  38}, 1913 (1988);
J. Main, G. Wiebusch, K. Welge, J. Shaw,
and J. B. Delos, Phys. Rev. A {\bf 49}, 847 (1994).

\bibitem{scaling}
This technique can be generalized to non-scaling systems by
treating $\hbar$ as a variable parameter and computing the
spectrum of $1/\hbar$ at constant $E$.  The Fourier transform
of the $1/\hbar$ spectrum gives peaks at the action of periodic
orbits and whose heights correspond to $A_{\it nk}$ in 
Eq. \ref{eqn:trace}.

\bibitem{simplify}  One might object that inserting
classical information at this point invalidates the
claim of being able to determine the initial angles
from the spectra.  The relationship between the initial
and final angles simplifies the discussion, but it
is not necessary for determining the initial conditions.
A more general technique which does not rely on this
relationship will be published elsewhere.

\bibitem{normalized}  This approach also works
by comparing absolute recurrence amplitudes
with absolute angular distributions.  However, absolute
recurrence amplitudes are prohibitively difficult to
measure experimentally.  Normalizing the angular
distributions and recurrence amplitudes circumvents this
difficulty and provides a technique for determining
the initial angles of closed orbits experimentally.

\bibitem{wintgen2} D. Wintgen and H. Friedrich, Phys. Rev. A
{\bf 35}, 1464 (1987).

\bibitem{clark} C.W. Clark and K.T. Taylor, J. Phys. B. {\bf 15}, 1175
(1982).

\bibitem{zimmerman}
M.L. Zimmerman, J.C. Castro, and D. Kleppner, Phys. Rev. Lett.
{\bf  40}, 1083 (1978).

\bibitem{van} T. van der Veldt, W. Vassen, and W.
Hogervorst, Europhys.  Lett.  {\bf 21}, 9 (1993).

\bibitem{shaw2}
Note the small peaks corresponding to the first few repetitions
of the perpendicular orbit.  
Closed-orbit theory predicts zero amplitude for this orbit.
($P_1(\cos{\theta}) = \cos{\theta}$, and $\theta = \pi/2$.)
This is because closed-orbit theory makes the assumption
that the recurrence amplitude carried by the neighboring trajectories
has the same initial angle as the closed orbit.
This approximation fails when the angular dependence of the
closed orbit is exactly zero, as in this case.
John Shaw and John Delos have modified closed-orbit
theory to account for the contribution of neighboring orbits
whose initial angles do not give a zero in the angular
dependence of the recurrence amplitude (private
communication).

\bibitem{baranger} M. Baranger, K. T. R. Davies, and J. H. Mahoney,
Ann. Phys. {\bf 186}, 95 (1988).
 
\end{thebibliography}
\end{document}